\documentclass[%
 aip,
 amsmath,amssymb,
 reprint,%
]{revtex4-1}

\usepackage{graphicx}% Include figure files
\usepackage{dcolumn}% Align table columns on decimal point
\usepackage{bm}% bold math
\usepackage[utf8]{inputenc}
\usepackage[T1]{fontenc}
\usepackage{mathptmx}
\usepackage{etoolbox}
\usepackage{mathtools}

\usepackage{soul}
\usepackage{xcolor}
\makeatletter
\def\@email#1#2{%
 \endgroup
 \patchcmd{\titleblock@produce}
  {\frontmatter@RRAPformat}
  {\frontmatter@RRAPformat{\produce@RRAP{*#1\href{mailto:#2}{#2}}}\frontmatter@RRAPformat}
  {}{}
}%
\makeatother
\begin{document}

\preprint{AIP/123-QED}

\title[Permanent magnet based Zeeman slower]{Permanent magnet based Zeeman slower for lithium atoms}
% Force line breaks with \\
\author{R. Elbaz}
\affiliation{ 
Department of Physics and QUEST Center and Institute of Nanotechnology and Advanced Materials, Bar-Ilan University, Ramat-Gan 5290002, Israel
}%
\author{F. Hamodi-Gzal}
\affiliation{ 
Department of Physics and QUEST Center and Institute of Nanotechnology and Advanced Materials, Bar-Ilan University, Ramat-Gan 5290002, Israel
}%
\author{N. Priel}
\affiliation{ 
Department of Physics and QUEST Center and Institute of Nanotechnology and Advanced Materials, Bar-Ilan University, Ramat-Gan 5290002, Israel
}%
\author{M. O.  Gzal}
% \homepage{http://www.Second.institution.edu/~Charlie.Author.}
\affiliation{%
Faculty of Mechanical Engineering, Technion Israel Institute of Technology, Haifa 3200003, Israel
}%
\author{L. Khaykovich}
\affiliation{ 
Department of Physics and QUEST Center and Institute of Nanotechnology and Advanced Materials, Bar-Ilan University, Ramat-Gan 5290002, Israel
}%
%\author{B. Author}%
% \email{Second.Author@institution.edu.}
%\affiliation{ 
%Authors' institution and/or address%\\This line break forced with \textbackslash\textbackslash
%}%

% \homepage{http://www.Second.institution.edu/~Charlie.Author.}
%\affiliation{%
%Second institution and/or address%\\This line break forced% with \\
%}%

\date{\today}% It is always \today, today,

\begin{abstract}
We describe the design, construction, and characterization of a permanent magnet based, transverse-field Zeeman slower for lithium atoms. We use off-the-shelf compact permanent bar magnets in the Halbach configuration to create a uniform magnetic field in the transverse direction. We develop a general approach for a mechanical structure that supports the spatial distribution of magnets using 3D printing technology. The approach allows for flexible assembly and dismantling of the magnetic field on the target vacuum system. Finally, we verify that the Zeeman slower supports a high flux of slow atoms in the region of magneto-optical trap. 
\end{abstract}

\maketitle

%\begin{quotation}
%The ``lead paragraph'' is encapsulated with the \LaTeX\ 
%\verb+quotation+ environment and is formatted as a single paragraph before the first section heading. 
%(The \verb+quotation+ environment reverts to its usual meaning after the first sectioning command.) 
%Note that numbered references are allowed in the lead paragraph.
%
%The lead paragraph will only be found in an article being prepared for the journal \textit{Chaos}.
%\end{quotation}

\section{\label{sec:intro}Introduction}

Most modern experiments with ultracold atoms start with deceleration of atoms that originates from an oven to below the capture velocity of the magneto-optical trap (MOT) which constitutes the next stage in the cooling process. 
Two main approaches to this initial stage are Zeeman slower~\cite{ZeemanPhillips82} and two-dimensional MOT (2D MOT)~\cite{2DMOTDieckmann98} and both methods are extensively used to decelerate a large variety of atoms and small molecules.
The efficient deceleration stage usually improves the overall performance of the experimental system, but in some cases it becomes essential.
Specifically, the initial slowing of lithium atoms is unavoidable due to their light mass and low vapor pressure at room temperature~\cite{LiApparatusHulet20}.

Both deceleration approaches are used intensively to provide a reliable beam of slow lithium atoms.
The advantages of 2D MOT include compact and simple design and the ability to work with low oven temperatures~\cite{2DMOTTiecke09,ModularPlatformHammel25}. 
Notable disadvantages, however, include a high-power laser and a large spatial extension of laser beams that are required to reach a flux of slow atoms comparable to that of Zeeman slower~\cite{ModularPlatformHammel25}.

More conventional and widely used sources of slow atoms are Zeeman slowers.
The traditional design uses a spatially varying longitudinal magnetic field combined with a circularly polarized, counter-propagating laser beam, resonant with $J\rightarrow J+1$ closed atomic transition~\cite{metcalf99}. 
By optimally tuning the Zeeman shift of the atomic energy levels, the resonance condition is maintained throughout the slowing process, continuously exerting a radiation pressure force that efficiently decelerates the atoms~\cite{OptimumDesignDedman04,Ohayon13,Ohayon15}. 
Typically, Zeeman slowers utilize electromagnets, requiring significant electrical currents to generate the necessary magnetic fields and allowing for increased capture velocity and a significant overall flux of slow atoms. 
These high-current systems commonly require water cooling, which adds experimental complexity, maintenance challenges, and potential stability issues. 
To resolve this issue to some extent, a zero-crossing magnetic field profile is used, where the field flips its sign along the trajectory, thus reducing the maximal absolute field needed. 
However, this does not solve the problem entirely.

In recent years, there has been significant interest in developing Zeeman slower designs based on permanent magnets~\cite{Ovchinnikov08,halbachZeemanRb11}. 
These offer substantial advantages by eliminating high-current demands and the associated cooling infrastructure. 
This approach results in simpler, more robust, and compact experimental setups that reduce complexity and facilitate reliable, maintenance-free operation.
The high flexibility of the device led to the realization of a dynamically configurable and optimizable structure of the Zeeman slower~\cite{ConfigurableZSReinaudi12}.

Generally, the magnetic field in permanent magnet based Zeeman slowers can be oriented either longitudinally or transversely with respect to the atoms' propagation axis. 
Although the more traditional longitudinal orientation suffers from a significant leakage of the field outside the slower region, it allows efficient use of circularly polarized light and realization of the zero-crossing configuration of the slower~\cite{LongitudinatOvchinnikov12,LongitudinalPermanentKrzyzewski14}.
A more elegant solution is to use the transverse-field configuration of permanent magnets, which creates a uniform field trapped mainly within the device, similar to the electromagnetic solenoid that creates a longitudinal field inside it~\cite{halbachZeemanRb11}. 
A drawback of this solution is the loss of 50$\%$ of laser power, related to the unavoidable superposition state of circular polarizations with respect to the quantization axis defined by the magnetic field.
A comparison between these two realizations for Sr atoms can be found in Ref.~\onlinecite{ZSforStrontiumHill14}. 

A notable disadvantage in using permanent magnets for a Zeeman slower arises from the strong mutual interactions between them.
The mechanical structure dedicated to maintain the magnets in the required spatial order is usually clumsy and massive, even if the number of magnets used in the contraction is kept to a minimum.~\cite{halbachZeemanRb11,LongitudinatOvchinnikov12,ConfigurableZSReinaudi12,LongitudinalPermanentKrzyzewski14,ZSforStrontiumHill14}.
In an early work, 3D printing technology has been used to create a mechanical support of permanent magnets for a Zeeman slower, although for a significantly smaller installation~\cite{3DPrintedZSParsagian15}.

%Additionally, certain permanent-magnet Zeeman slower designs employ a transverse magnetic field orientation rather than the conventional longitudinal configuration. 
%This choice often facilitates achieving a near-zero residual field outside the slower region, which can be advantageous for subsequent atom trapping and manipulation.

In this paper, we present a transverse-field Zeeman slower designed for lithium-7 atoms. 
Our device utilizes a Halbach array composed of small off-the-shelf bar magnets integrated into a 3D-printed structure which hosts 136 of them. 
Despite the large number of magnets, the implementation remains notably simple, flexible, cost-effective, and mechanically robust, making it suitable for practical applications and reliable long-term operation.
In addition, the 3D-printed structure allows flexible assembly and dismantling of the magnetic field, allowing efficient backing of the vacuum system.
Previously, permanent magnets were used to create part of the magnetic field only in the so-called hybrid Zeeman slower configuration for lithium atoms~\cite{HybridZSGarwood22}.
Here, the entire device is constructed from permanent magnets, and a large flux of slow atoms is demonstrated in the region of the magneto-optical trap without a significant bias field.

\section{Design and construction}

\subsection{\label{sec:OpMech}Principles of operation}

\begin{figure}
\includegraphics[width=\columnwidth]{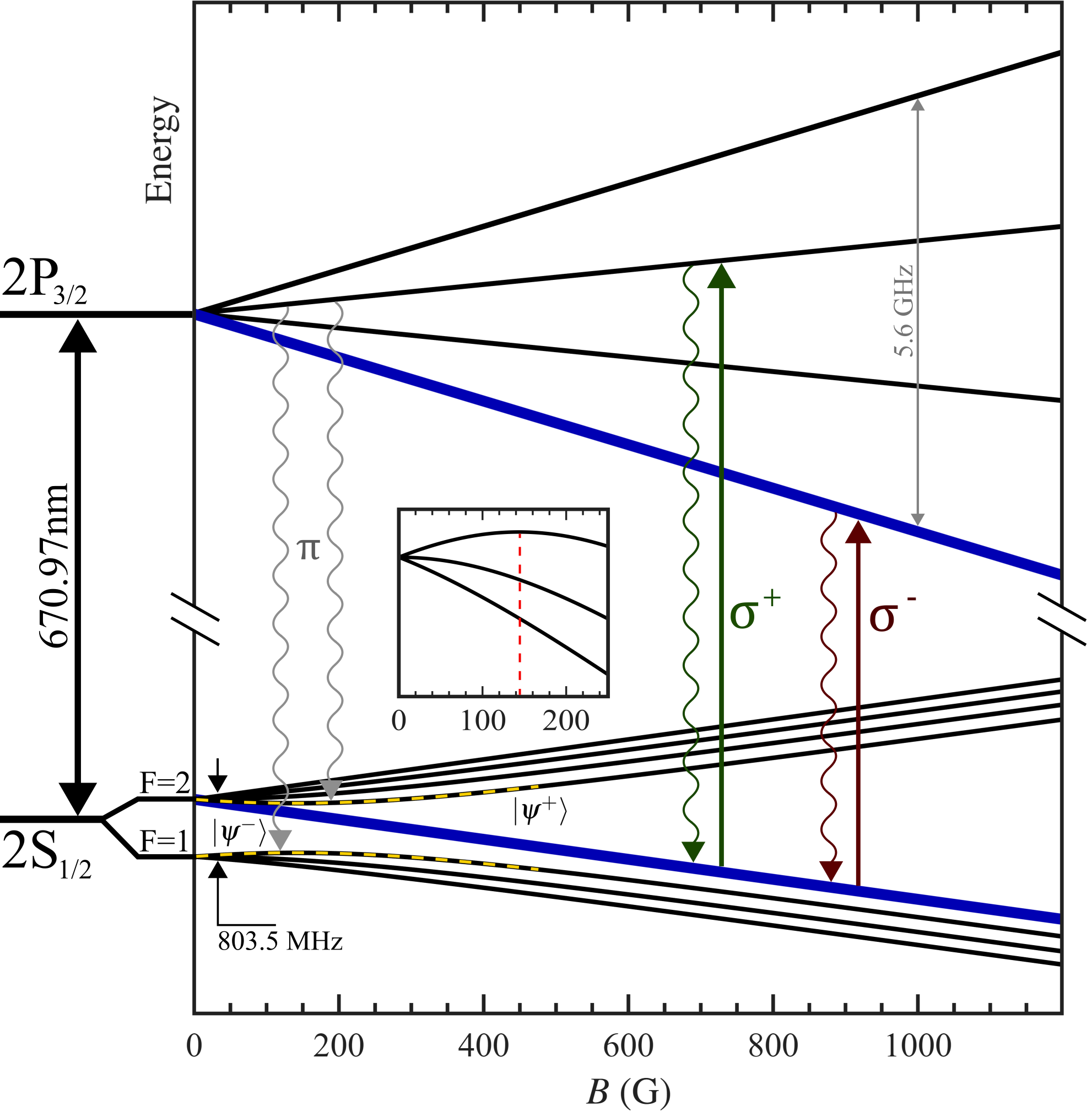}
\caption{\label{fig:energyDiagram} Relevant energy spectrum of $^7$Li as a function of magnetic field. The two states used for the slowing mechanism are shown as thick blue lines. The primary excitations from our target ground state are indicated by straight arrows for $\sigma^+$/$\sigma^-$ polarized light. Spontaneous emission paths from these two excited states are represented by the wavy lines. The two possible ground states reached via a $\sigma^+\rightarrow\pi$ cycle are highlighted with dashed yellow lines. The inset zooms into the $F=1$ ground-state manifold and marks the transition from $m_F$ to $m_J^\prime$ states at $B_{turn}=143$ G (dashed red line). Apart from the illustrative axis break between the $2S_{1/2}$ and $2P_{3/2}$ manifolds, all data in the main figure is to scale (the hyperfine splitting in the $2P_{3/2}$ state is of the order of a few MHz and therefore not visible here.)}
\end{figure}

The relevant energy-level scheme of the $^7$Li atom subject to a magnetic field is shown in Fig.~\ref{fig:energyDiagram}.
To induce the slowing process, the laser light addresses a pair of energy levels that are indicated by thick blue solid lines.
%Note that we choose the increasing magnetic field configuration.
 
The maximal achievable deceleration exerted on a two-level atom by a counter-propagating resonant laser beam is given by
\begin{equation}
    \label{aMax}
    a_{max}=\frac{\hbar k\Gamma}{2m},
\end{equation}
where $k=2\pi/\lambda$ is the wavevector, $\Gamma$ is the linewidth of the atomic transition, and $m$ is the atomic mass~\cite{metcalf99}.
%We choose to work with an increased magnetic field Zeeman slower configuration.
To sustain resonance conditions throughout the slowing region, the Zeeman slower's magnetic field profile must satisfy:
\begin{equation}
    \label{resonanceCond}
    \delta_0 + kv(z) + \frac{\mu}{\hbar} B(z) = 0,
\end{equation}
where $\delta_0 = \omega_L - \omega_0<0$ is the laser detuning from the atomic transition frequency $\omega_0$ at a zero magnetic field, $v(z)$ is the atomic velocity at position $z$, and $\mu = h \times 1.4\;\text{MHz/G}$ represents the magnetic moment difference relevant to our system. 
To accommodate practical deviations from the ideal magnetic field profile and finite laser intensity, a tolerance factor $\eta<1$ is introduced so that the target deceleration is set to $a = \eta a_{max}$. 
A lower value of $\eta$ increases the tolerance, but extends the required length of the slower. 
In our design, we selected a moderate tolerance with $\eta = 0.5$.

Given an initial capture velocity, the required magnetic field profile along the slower can be expressed analytically as:
\begin{equation}
    \label{BofZ}
    B(z) = B_0 - B_c \sqrt{1 - z/L} \quad \text{for} \quad 0<z<L,
\end{equation}
where $B_0 = \hbar |\delta_0|/\mu$ corresponds to the resonant field at zero atomic velocity, $B_c = \hbar k v_c/\mu$ characterizes the magnetic field range required to slow down atoms from the so-called capture velocity $v_c$ down to zero velocity, and $L=v_c^2/2a$ is the length required for that process. 
The dashed red curve in Fig.~\ref{fig:magnets}(a) shows the ideal field profile for our system.
Note that we have chosen the increasing magnetic field configuration.

Due to the transverse magnetic field orientation, the slowing laser's linear polarization must be perpendicular to the magnetic field, corresponding to a superposition of circularly polarized components: $\pi_0 = \sigma^+ + \sigma^-$. 
This indicates that only 50\% of the light, corresponding to the $\sigma^-$ polarization, drives the desired closed transition (see Fig.~\ref{fig:energyDiagram}), while the remaining $\sigma^+$ component can induce unwanted excitations and optically pump the atoms into dark states.
This process is maximal at low field values, where the Zeeman shift of the transitions corresponding to the polarization components $\sigma^+$ and $\sigma^-$ are nearly degenerate.
Thus, in order to eliminate this problem, it is beneficial to start the deceleration from a nonzero initial magnetic field that Zeeman shifts the unwanted transitions governed by $\sigma^+$ polarization. 
%This suppresses unwanted optical pumping of atoms at the target velocity group, as the detuning of the transition driven by $\sigma^+$ polarization becomes large throughout most of the slowing trajectory.

Initially, at the output of the oven, the atoms are equally distributed among eight energy levels of two hyperfine ground states (see Fig.~\ref{fig:energyDiagram}), while the deceleration addresses only a single energy level.
In order to avoid the loss of $87.5\%$ of the population we perform initial optical pumping of atoms by introducing a laser light perpendicular to the atomic beam that is purely $\sigma^-$ polarized with respect to the relevant quantization axis.
The optical pumping region is prior to the Zeeman slower region.

%Naturally, for longitudenal field configuration, one can add a repumper frequency to the slowing beam and get optical pumping into the slowing state due to pure $\sigma^-$ polarization. 
%Here, another elegant solution is made possible thanks to the residual field prior to the slowing region. 
%A beam perpendicular to the atomic beam can have pure $\sigma^-$ polarization with respect to the relevant quantization axis. 
%This allows for a simple single frequency slowing beam.

\subsection{\label{sec:Construct}Design considerations}

\subsubsection{\label{sec:mechStruct}Mechanical structure}

\begin{figure}
\includegraphics[width=\columnwidth]{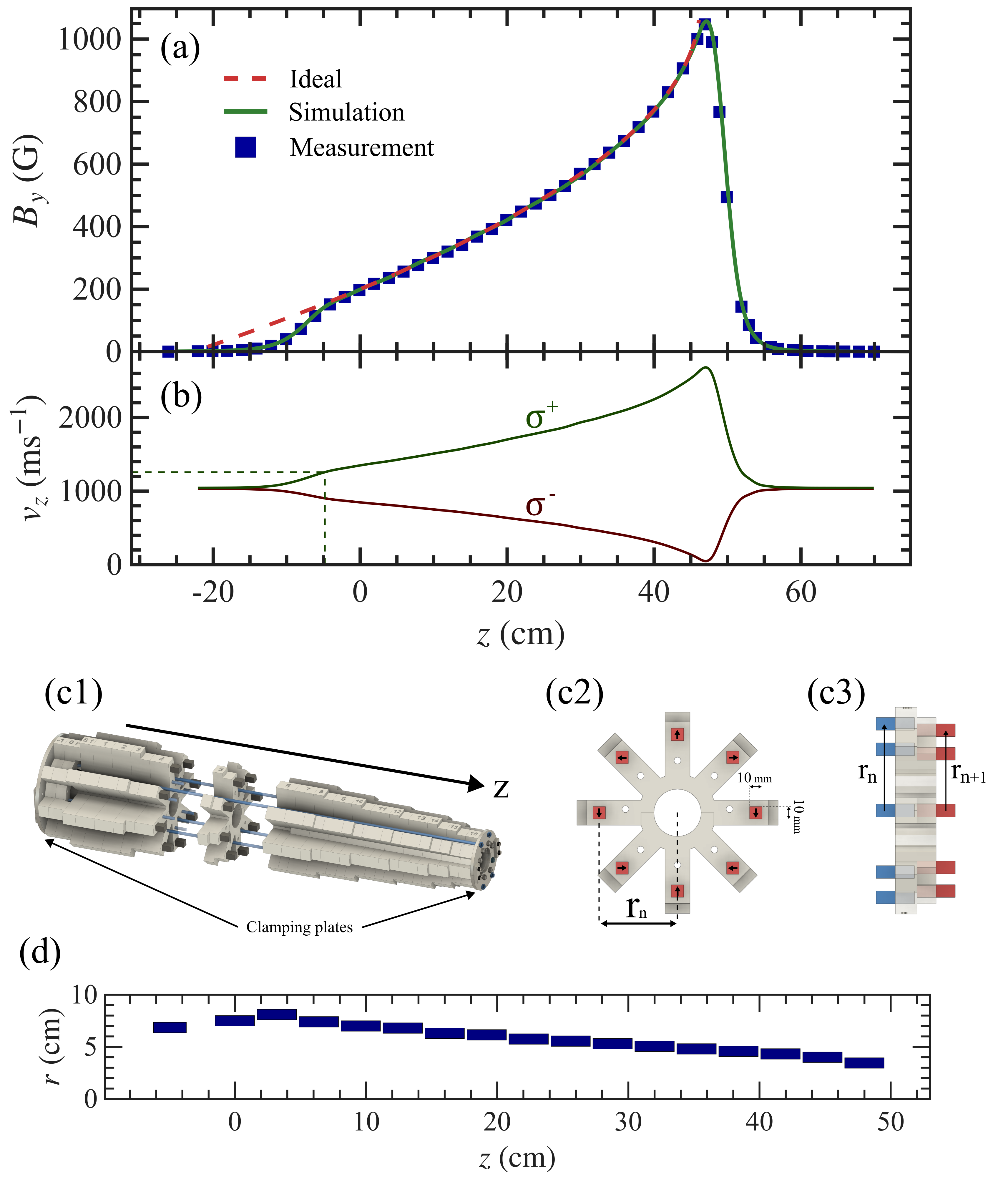}
\caption{\label{fig:magnets} Construction of the magnetic field profile: (a) ideal (dashed red line), calculated (solid green line) and measured (blue dotes) $\hat{y}$-component of the magnetic field profile along propagation axis of the Zeeman slower; (b) Velocity in resonance with the slowing laser along the Zeeman slower. The dashed lines mark the location at which the field crosses $B_{turn}$ and the corresponding velocity that interacts with the $\sigma^+$ component; (c1)-(c3) show the 3D printed mechanical structure that holds permanent magnets. The black arrows on the magnets in (c2) mark the magnetization axes; (d) shows to scale bar magnets position and orientation in a single Halbach leg of the Zeeman slower.}
\end{figure}

\begin{figure}
    \includegraphics[width=\linewidth]{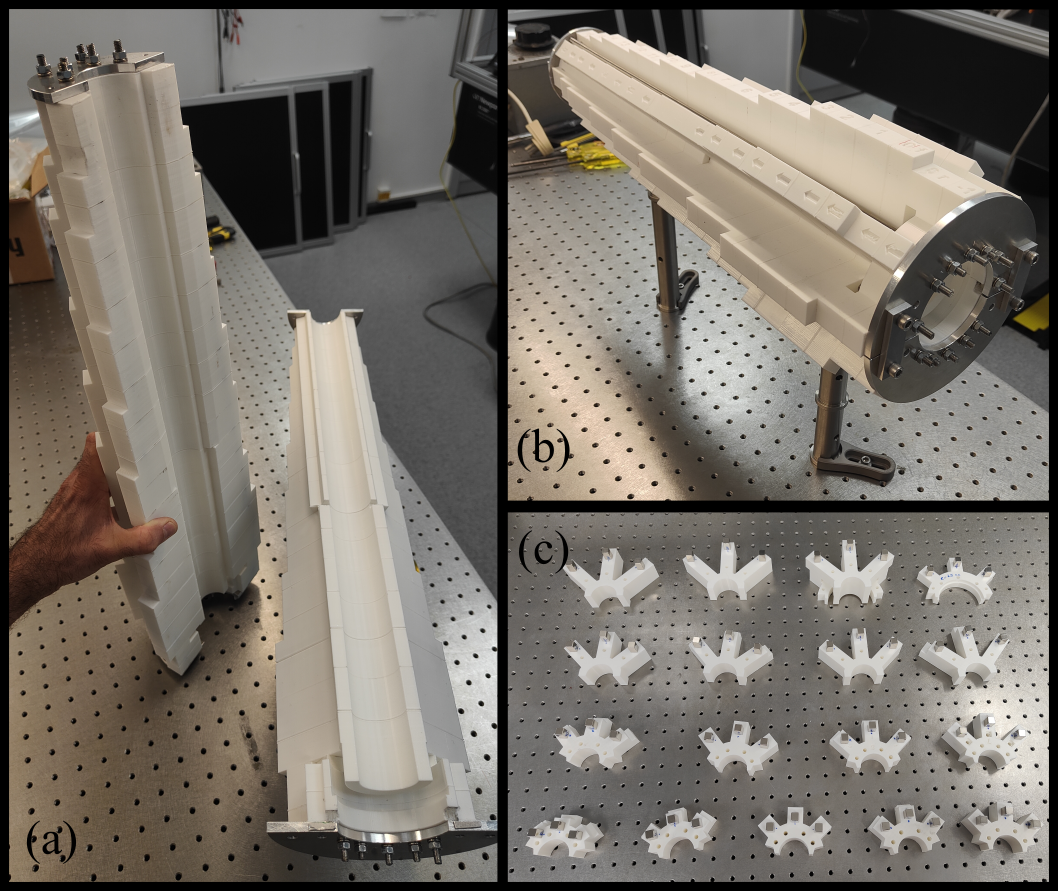}
    \caption{\label{fig:photos} Selected pictures of the Zeeman slower. (a) The top and bottom parts disassembled. (b) The complete structure, mounted at the designated height. (c) Separated layers of the bottom part showing the internal structure holding the magnets in place.}
\end{figure}

To experimentally realize the desired magnetic field profile, we design and construct a sequence of cylindrical Halbach arrays aligned along the $z$-axis (see Fig.~\ref{fig:magnets}(c1)). 
Each array consists of eight NdFeB bar magnets (HKCM Magnet-Cuboid Q30x10x10Ni-N42 and Q25x10x10Ni-N35) arranged in a Halbach configuration (Figs.~\ref{fig:magnets}(c2) \&~\ref{fig:photos}(c)), generating a radially uniform and predominantly transverse magnetic field direction ($\hat{y}$) along the atomic trajectory.
Each array is characterized by a radius $r_n$ and is positioned at discrete locations $z_n$ along the slowing region, allowing fine control over the local field amplitude. 
Fig.~\ref{fig:magnets}(d) shows the position of the magnets along the device.

The magnets are housed within a 3D printed support structure, where each slice holds half of the magnets of two adjacent Halbach arrays (Fig.~\ref{fig:magnets}(c3)). 
This interleaved design results in a rigid assembly that can withstand the significant magnetic forces between individual magnets.
A standard PLA material is used for all 3D printed parts.
To simplify installation on the UHV system, the structure is divided into two separate parts: top and bottom, which can be independently mounted and secured (see Fig.~\ref{fig:photos}(a)).

Structural reinforcement is provided by stainless steel rods inserted through dedicated channels in the assembly. 
These rods are fastened with bolts at both ends, counteracting the repulsive forces between the magnet plates. 
To prevent deformation or damage to the 3D printed components, aluminum plates are placed at both ends of the device, providing robust metal interfaces for the bolts and ensuring reliable mechanical stability (see Fig.~\ref{fig:photos}(b)).

The design we use can be realized with a smaller number of longer magnets, which might decrease the mechanical complexity of the structure.
However, a large number of small magnets allows maximal flexibility in the realization of magnetic field profile and uniformity of the mechanical structure (see Section~\ref{sec:magConfig}).
The internal structure of lithium atoms makes the transfer field Zeeman slower with spin-flip configuration impractical~\cite{LongitudinatOvchinnikov12}.
However, for other atomic species (Sr and Yb) such a configuration is desirable, and small off-the-shelf magnets can provide a highly flexible solution in constructing a suitable magnetic field profile.

\subsubsection{\label{sec:magConfig}Magnets configuration}

The field generated by a single Halbach array is calculated following the methodology described in Ref.~\onlinecite{halbachZeemanRb11}. 
For simplicity and mechanical uniformity, all magnets are oriented with their long axis aligned along the $z$-direction. 
Although the finite size of the magnets introduces small oscillatory deviations from the ideal profile, the simplicity of this design was prioritized, and as we demonstrate below, it does not compromise performance.
The positions of the magnets are manually optimized to approximate the ideal magnetic field profile as closely as possible while accommodating the geometric constraints of our experimental apparatus. 

Ideally, the magnetic field should follow Eq.~(\ref{BofZ}), decreasing sharply to zero at both ends of the slowing region, with the spatial boundaries defined by the capture velocity $v_c$ and the final velocity $v_f$. 
However, in practice, implementing such a profile with discrete magnetic elements leads to an extension of the field outside the sharp boundaries.
%In addition, unavoidable field modulation between the adjacent magnets distorts the magnetic field profile. 
This can decelerate atoms beyond the intended final velocity, occasionally reaching near-zero or even negative velocities. 
Overslowing results in increased transverse diffusion and beam divergence due to prolonged interaction times. 
To avoid this, we intentionally relaxed the magnetic field gradient near the end of the slower region, ensuring a well-defined final velocity set by the maximum field.

The magnetic field component along the $\hat{y}$-direction was measured along the $\hat{z}$-axis using a Gaussmeter. 
A comparison between the ideal, calculated, and measured fields is shown in Fig.~\ref{fig:magnets}(a). 
The slowing region is set to begin at $z=0$ with an initial magnetic field of $B_i = 200$~G, and a capture velocity of $v_c = 850$~m/s. 
In Section~\ref{sec:captField}, we show that the measured field follows the ideal trajectory sufficiently beyond the $z=0$ limit, allowing atoms with velocities above that value to still experience deceleration within the slower, thereby effectively increasing the system's capture velocity.

A residual field of $\sim1.2$ G is measured in the center of the MOT, located $12$~cm behind the clamping plate. 
%(THIS MIGHT NOT BE THE BEST PLACE FOR THIS SENTENCE)

\section{Methods}

\subsection{\label{sec:measurementApparatus}Experimental setup}

\begin{figure*}
\includegraphics[width=\textwidth]{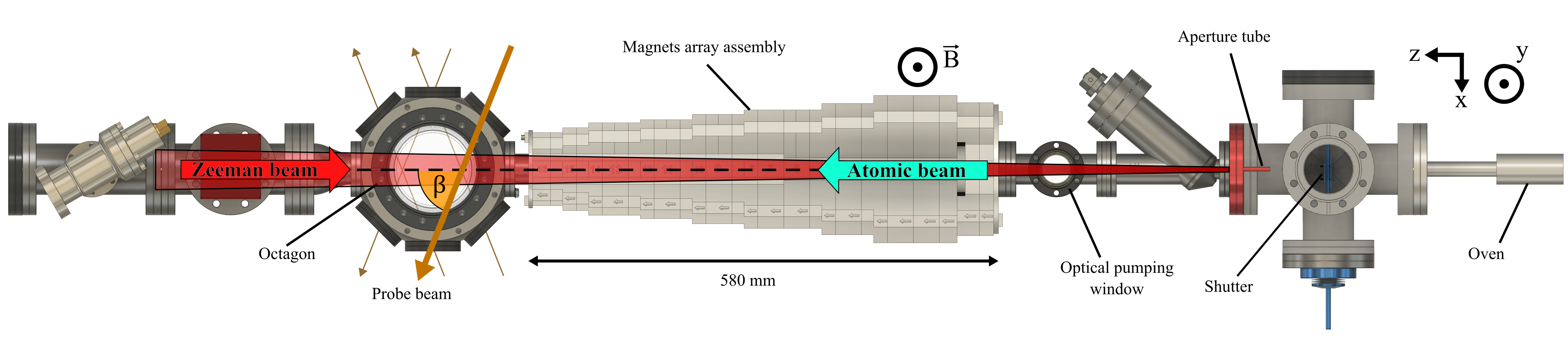}
\caption{\label{fig:system}Schematic representation of the experimental apparatus to demonstrate the performance of Zeeman slower. Atomic beam (light blue large arrow) originates from oven and propagates from right to left. It is collimated by the aperture tube and crosses the optical pumping region before entering the Zeeman slower deceleration region. The slowing laser beam (red large arrow) propagates from left to right and the probe laser beam is marked by a dark orange arrow which makes an angle $\beta$ with the atomic beam.}
\end{figure*}

A schematic representation of the experimental setup is shown in Fig.~\ref{fig:system}. 
$^7$Li atomic vapor is generated in an oven held at different temperatures $T$ and emitted through an aperture tube (4 mm in diameter, 50 mm in length), forming a collimated atomic beam in the $\hat{z}$ direction. 
A mechanical shutter positioned upstream of the aperture allows the atomic beam to be blocked when needed. 
Before entering the Zeeman slower, atoms are optically pumped into the desired slowing state using a pure $\sigma^-$ polarized beam (see Section~\ref{sec:opticalPumping}).

% \begin{figure*}
% \includegraphics[width=\textwidth]{system.png}
% \caption{\label{fig:system}Schematic representation of the experimental apparatus to demonstrate the performance of Zeeman slower. Atomic beam (light blue large arrow) originates from oven and propagates from right to left. It is collimated by the aperture tube and crosses the optical pumping region before entering the Zeeman slower deceleration region. The slowing laser beam (red large arrow) propagates from left to right and the probe laser beam is marked by a dark orange arrow which makes an angle $\beta$ with the atomic beam.}
% \end{figure*}

The slowing region extends from $z = 0$ to approximately $z \approx 47$ cm, corresponding to a capture velocity of $v_c = 850$ m/s and a final velocity of $v_f = 46$ m/s. 
In practice, the effective slowing region extends slightly beyond these boundaries due to laser power broadening and the gradual onset and fall-off of the magnetic field. After exiting the slower, the atoms enter the science chamber (octagon), where they are captured into the MOT.
The slowing beam counter-propagates along the $\hat{z}$ axis, entering through a $ 4.5''$ viewport at the far end of the apparatus. 
It is focused to match the expected transverse spread of the atomic beam, with a measured waist of $w_2 = 5\;\text{mm}$ near the oven-side tube entrance and $w_1 = 11.5\;\text{mm}$ at the center of the octagon.

All lasers originate from a tunable diode laser (DL pro, Toptica Photonics), locked to the cross-over feature of the $^7$Li D2 transition.
The optical pumping frequencies are generated by shifting the light by $\pm401.75\;\text{MHz}$ using a double-pass 200 MHz AOM (G\&H 3200-125). 
For characterization of the optical pumping (OP) process, both frequencies are injected into a tapered amplifier (Toptica EYP-TPA-0670), which yields a combined optical power of up to $P_{OP}=60\;\text{mW}$ on the atoms. 
A key advantage of the transverse magnetic field geometry is that the optical pumping can be performed with a beam perpendicular to the atomic velocity, thereby uniformly addressing the entire velocity distribution due to negligible Doppler broadening.

The slowing beam is generated by shifting the laser frequency by $-1948\;\text{MHz}$ relative to the locking point, using a sequence of three 200 MHz AOMs (G\&H AOMO 3200-125) and one 350 MHz AOM (AA Optoelectronic MT350-A0.12-800) all in double pass configuration. 
The slowing beam is amplified by a separate tapered amplifier, which provides up to $P_z \approx 180\;\text{mW}$ of power at the atoms, although only about 50\% of the light is in the desired polarization state.

\subsection{\label{sec:Meas}Measurement methodology}

To characterize the Zeeman slower performance we perform Doppler-sensitive absorption spectroscopy on atoms flying across the MOT region. 
A probe beam (waist $w_p=2.3$ mm, intensity $I_p = 0.2 I_{\text{sat}}$) is directed through the atomic beam in the midplane ($y = 0$), forming an angle $\beta = 68^\circ$ with respect to the $x$ axis (marked as an orange arrow in Fig.~\ref{fig:system}). 
The $z$-component of the probe's wavevector is aligned with the atomic beam direction. 
The probe passes through the octagon four times, maintaining a nearly constant angle with a deviation of at most $\Delta\beta \approx 0.5^\circ$.

We measure the transmitted power of the probe beam with the atomic shutter both open and closed ($P_{open}$ and $P_{closed}$ respectively), while scanning the probe frequency $f$ in the vicinity of the D2 spectroscopy line. 
The frequency is calibrated on a Li vapor cell using Doppler-free spectroscopy. 
From these 2 measurements, we extract the frequency-dependent optical density, $\text{OD}(f)=\log\left(P_{closed}/P_{open}\right)$. 
In order to link this measurement with the atomic velocity we project the observed Doppler shift on the $z$ axis:
\begin{equation}
    v_z=\frac{\Delta_2}{k\cos(\beta)}
\end{equation}
where $\Delta_2$ is the probe detuning from the atom transition at $F=2$ at zero magnetic field. 
%Of course, this formula is not valid for $\beta\rightarrow90\degree$ where one needs to take into account the finite spread of transverse velocities (so-called $\sigma_\bot$). %For almost all our measurements, we can safely rely on the fact that $\sigma_\bot \sin(\beta)\ll\sigma_z \cos(\beta) \; , \; v_z$, ({\lk Check this please!! I think it should be $\sigma_\bot \sin(\beta)\ll v_z \cos(\beta)$}) and therefore does not change our results significantly.

The measured OD of atoms slowed by the Zeeman slower is the result of a convolution between the velocity-dependent atomic density distribution (assumed to be spatially uniform):

\begin{equation}
    \label{density}
    n(v_\beta)=\frac{n_0}{\sqrt{2\pi\sigma_\beta^2}}e^{-\:\frac{(v_\beta-\bar{v}_\beta)^2}{2\sigma_\beta^2}}
\end{equation}
and the velocity- and detuning-dependent scattering cross-section:
\begin{equation}
    \label{crossSection}
    \sigma(v_\beta, \Delta_2)=\frac{\sigma_0}{1+s_0+4\left(\frac{\Delta_2-kv_\beta}{\Gamma}\right)^2},
\end{equation}
where $n_0$ is the peak atomic density, $v_\beta$ is the velocity along the probe's wavevector axis, $\bar{v}_\beta$ and $\sigma_\beta$ are the mean and standard deviation of the normal distribution of velocities, $\Gamma=2\pi\times5.9$~MHz is the linewidth of the atomic transition, $\sigma_0$ is the on-resonance cross-section, and $s_0=I_p/I_{sat}$ is the saturation parameter.

The resulting OD is then given by the following:

\begin{equation}
\label{OD}
\text{OD}(\Delta_2) = \frac{\pi \; l \; n_0 \; \sigma_0 \; \gamma}{1 + s_0} \cdot V(\Delta_2/k-\bar{v}_\beta; \gamma, \sigma_\beta),
\end{equation}
where $l$ is the effective path length, $\gamma=\frac{\Gamma\sqrt{1+s_0}}{2k}$, and $V$ denotes the normalized Voigt profile.
By fitting this function with the measured OD for velocities in the range $-200<v_z<200$ m/s, with $\bar{v}_\beta$ and $\sigma_\beta$ as fitting parameters, we can extract the atomic density $n_0$. This allows for the estimation of the flux of slow atoms passing through the MOT region:

\begin{equation}
    \label{flux}
    \Phi=n_0 \;A \; v_f
\end{equation}
where $A=\pi\cdot12$ mm$^2$ is the area defined by diameter of the MOT beams, and $v_f$ is the most probable final velocity along the $z$ axis extracted from the data.
%It is worth mentioning that this number is an underestimate of the actual flux. 
%Although multiple passes of the probe beam through the octagon improve our SNR, it also induces some redistribution of slow atoms among the $F=1$ states due to the lack of a clear quantization axis. 
%This can be seen with the naked eye in the results shown in Fig.~\ref{fig:velDistVaryDelta}. 
%However, a relatively good quantitative understanding can still be obtained by collecting data from atoms in the $F=2$ state, and %therefore we have decided not to further pursue this missing portion of the flux.
%{\lk I am a bit confused here. What can be seen with the naked eye?}

\section{\label{sec:VelDist}Results}
\subsection{\label{sec:SubVelDist}Velocity distribution}
\begin{figure*}
\includegraphics{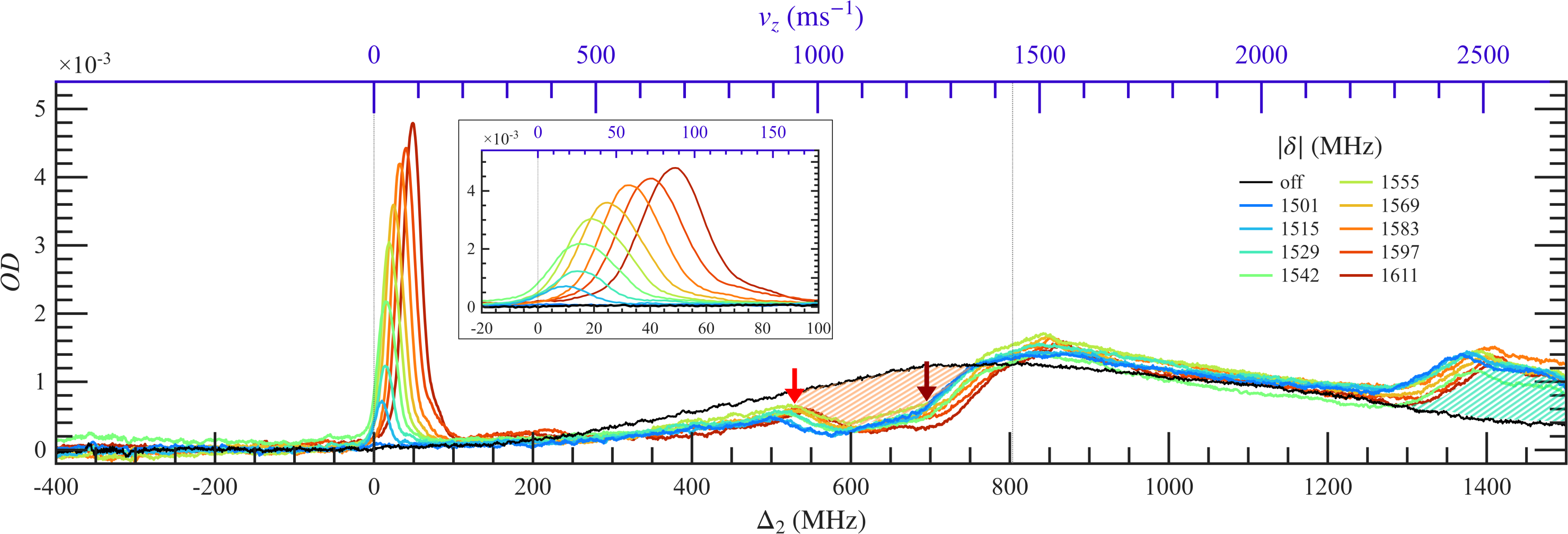}
\caption{\label{fig:velDistVaryDelta} The measured velocity profiles as a function of probe laser detuning (lower axis) and velocity (upper axis) for a number of slowing laser detunings. The initial Maxwell-Boltzmann distribution of velocities in the atomic beam is shown as a black solid line. The colored lines represent velocity profiles for different detunings as marked in the legend. The inset represent peaks of slow velocities. Vertical doted lines mark the ground state hyperfine splitting in lithium atoms. For the explanations of other features see text.}
\end{figure*}

We begin by characterizing the slowing efficiency for various values of the slowing beam detuning $\delta_0$. 
%Since the magnetic field along the device was measured prior to its installation on the vacuum system, some uncertainty remains regarding the actual field profile when the apparatus is assembled in place. 
%This uncertainty is highest near the peak of the magnetic field, which corresponds to the final velocity reached by the slowed atoms.
%Additionally, transverse atomic diffusion may lead atoms to explore regions of higher magnetic field at larger radial positions. 
%Together, these effects can induce some variation of the detuning for optimal performances.
For this measurement, the oven temperature is set to $T = 530^\circ$C, the slowing beam power is $P_z = 180$ mW, and the optical pumping is on with a total power of $P_{\mathrm{OP}} = 60$~mW. 

Fig.~\ref{fig:velDistVaryDelta} shows the optical density (OD) as a function of velocity (upper axis) and probe laser detuning $\Delta_2$ (lower axis) for various values of $\delta_0$.
The hyperfine splitting of the ground state of the $^7$Li atoms is marked by two gray dotted lines that are separated by $803.5$~MHz. 
$\Delta_2=0$ ($\Delta_2=803.5$~MHz) corresponds to the $F=2\rightarrow F^\prime$ ($F=1\rightarrow F^\prime$) transition, respectively. 
The reference measurement is taken with the slowing beam turned off while the optical pumping is kept on.
This measurement is represented as a black solid line in Fig.~\ref{fig:velDistVaryDelta} and shows a typical Maxwell-Boltzmann velocity distribution profile.
Note that OP cleans the $F=1$ state well, leading to a nearly perfect single-peak structure of the reference profile.

A prominent feature in the data when the slowing laser is on (all colored solid lines) is the accumulation of slow atoms with final velocities below $100$ m/s. 
The inset in Fig.~\ref{fig:velDistVaryDelta} highlights a gradual shift in the most probable final velocity $v_f$ toward lower values that follows the increase in detuning, as expected.

\begin{figure}[h]
\includegraphics{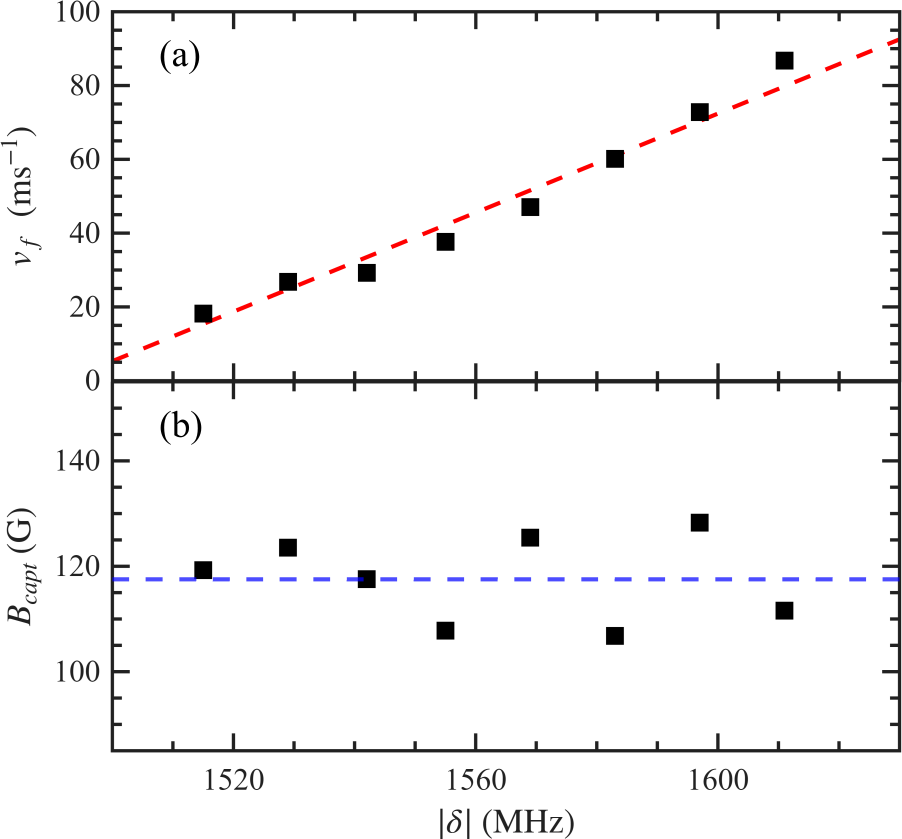}
\caption{\label{fig:finalVelCaptFieldVsDelta} (a) The most probable final velocity $v_f$ and (b) the extracted capture magnetic field $B_{capt}$ as a function of slowing laser detuning. The dashed red line in (a) corresponds to a linear function with the slope $\lambda = 670$~nm and no fitting parameters. The dashed blue line in (b) indicates the mean value of the measurements.}
\end{figure}

Fig.~\ref{fig:finalVelCaptFieldVsDelta}(a) shows the most probable final velocity $v_f$ as a function of $\delta_0$. 
As expected (according to Eq.~(\ref{resonanceCond})), the points follow a linear trend with slope $\lambda$ (red dashed line in figure). 
Small deviations from linearity appear at both ends of the detuning range. 
At small $|\delta_0|$, the atoms are slowed to a near-zero longitudinal velocity, leading to increased transverse spreading and a corresponding loss of signal. 
At large $|\delta_0|$, the atoms begin to outrun the local resonance condition and do not follow the slowing trajectory, resulting in a higher final velocity.

\subsection{\label{sec:captField}Capture magnetic field}
Due to the mixed polarization of the slowing light, two distinct processes occur depending on atomic velocity (see Fig.~\ref{fig:magnets}(b)). 
Atoms with $v_z < v_c$ become resonant with the slowing laser in regions where $B > B_i$ and undergo a closed cycle governed by $\sigma^-$ light. 
In these fields, the transitions induced by $\sigma^+$ polarization are significantly detuned, and thus suppressed. 
However, atoms with $v_z > v_c$ can remain resonant with the $\sigma^+$ component in the interaction region and eventually undergo a $\sigma^+$ absorption and $\pi$ spontaneous emission cycle, resulting in optical pumping to one of the states marked with dashed yellow lines in Fig.~\ref{fig:energyDiagram}, namely $\left| \psi^{+} \right>$ and $\left| \psi^{-} \right>$. 
%({\lk $\left| \psi^{+} \right>$ and $\left| \psi^{-} \right>$ are small in Figure}). 
At field values $B\gg B_{turn}$ (marked in red in the inset of Fig.~\ref{fig:energyDiagram}) this transition only populates the upper state $\left| \psi^{+} \right>$, which adiabatically traces back to the $F = 2$ ground state. 
However, in low fields, the atoms can decay to the bottom state $\left| \psi^{-} \right>$ due to state mixing induced by hyperfine interaction.
These atoms are depleted from the $F=2$ population (marked as a red slashed region in Fig.~\ref{fig:velDistVaryDelta}) and increase the $F=1$ population (light blue slashed region).

Therefore, there are two interesting thresholds in the velocity distribution. 
The first is the transition across $B_{turn}$ (marked with a brown arrow in Fig.~\ref{fig:velDistVaryDelta}) where optical pumping to $\left| \psi^{+} \right>$ becomes favorable over $\left| \psi^{-} \right>$. 
Not surprisingly, this occurs at a velocity of $v_z\approx1260$ m/s, which corresponds to $B_{turn}$ at our laser frequency for $\sigma^+$ light.
The second is the threshold between the optically pumped and effectively slowed atoms, and it is of greater importance for the characterization of our system. 
It is identified as the local maximum near the target $v_c$ (marked with a red arrow in Fig.~\ref{fig:velDistVaryDelta}).

%The strong depletion of atoms at higher velocities is attributed to depumping caused by the unwanted polarization component.

%To characterize this turning point more systematically, we define it as the velocity at which the scattering of $\sigma^-$ light starts to dominate over the optical pumping caused by $\sigma^+$ light. 
To characterize the last turning point more systematically, we extracted it from the smoothed velocity distributions as the local maximum near $v_z \approx 900$ m/s. 
Translating this velocity to its corresponding resonant magnetic field allows us to identify the capture field value ($B_{\mathrm{capt}}$) where the atoms effectively begin the slowing process.

Fig.~\ref{fig:finalVelCaptFieldVsDelta}(b) shows the extracted $B_{\mathrm{capt}}$ as a function of the laser detuning $|\delta_0|$.
Although the slower was designed to have an initial field around $200$~G, the observed capture field remains consistently lower, around 115-125 G, for all detunings. 
This indicates that atoms typically begin to interact with the slowing beam earlier than anticipated. 
The near-constant magnetic field $B_{\mathrm{capt}}$ across detunings suggests that the field profile follows the ideal trajectory well enough to allow robust capture despite slight mismatches or experimental uncertainties in the field or beam alignment.

\subsection{\label{sec:fluxVsPz} Slowing beam power}

We further examine the performance of the slower as a function of the slowing beam power $P_z$, for several values of laser detuning $\delta_0$. 
All measurements in this series are performed with the optical pumping on with a total power of $P_{\mathrm{OP}} = 60$ mW.

\begin{figure}
\includegraphics{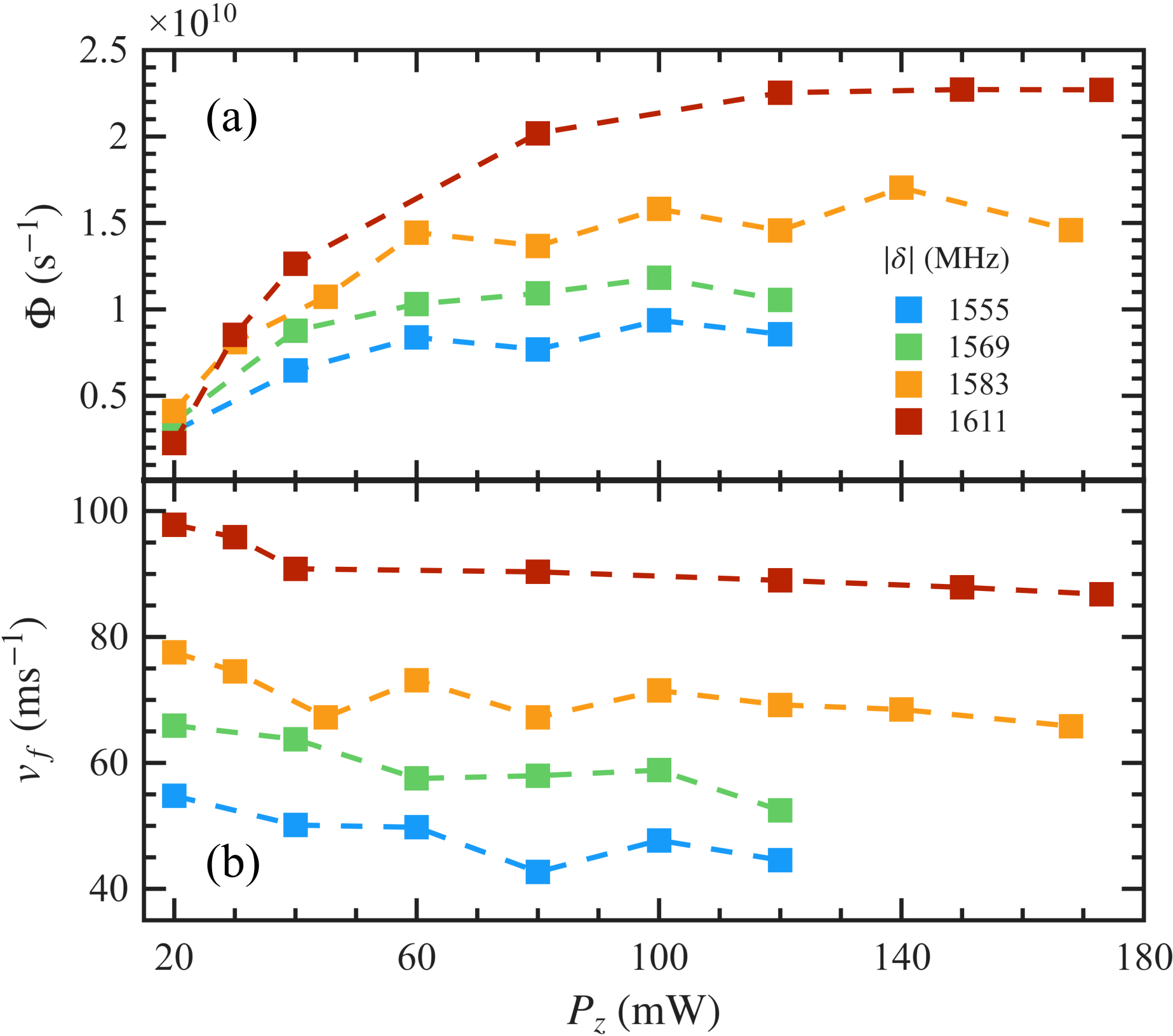}
\caption{\label{fig:fluxAndVelVsPz} (a) The flux of slow atoms $\Phi$ and (b) the the final most probable velocity $v_f$ as a function of slowing laser power foe different slower laser detunings.}
\end{figure}

Fig.~\ref{fig:fluxAndVelVsPz}(a) shows the measured flux $\Phi$ of slowed atoms, and panel (b) shows their corresponding final velocity $v_f$, both plotted against $P_z$ for four different values of $|\delta_0|$.

As $P_z$ increases, the slowing beam is able to maintain resonance with atoms across a wider range of trajectories, including those that deviate transversely from the central axis. 
This leads to a clear increase in flux with power, which eventually saturates as the transition is fully power-broadened. 
Additionally, for a fixed power, the flux increases with larger detunings (more negative $\delta_0$). 
This behavior is attributed to the higher capture velocities associated with a deeper detuning, allowing a larger portion of the initial thermal distribution to be slowed.

In Fig.~\ref{fig:fluxAndVelVsPz}(b), we observe that the final velocity $v_f$ tends to decrease as the slowing beam power increases. 
This behavior is attributed to the effect of power broadening, which allows atoms to remain resonant with the slowing light even beyond the point of maximum magnetic field, where the field no longer matches the ideal slowing trajectory for their velocity. 
As a result, the deceleration force remains significant over a longer distance, enabling the atoms to be slowed further.
% than they otherwise would.

\subsection{\label{sec:opticalPumping}Optical Pumping}

In this section, we investigate the dependence of the flux of slow atoms on optical pumping.

The optical pumping is performed just before the atoms enter the slower region (see Fig.~\ref{fig:system}).
The quantization axis for this process is well defined by the residual magnetic field at the entrance of the slower.
The optical pumping beam contains both the pump and the repump frequencies, referenced to the zero-field atomic transitions, and its waist is set to $w \approx 3\;\text{mm}$.

\begin{figure}
\includegraphics{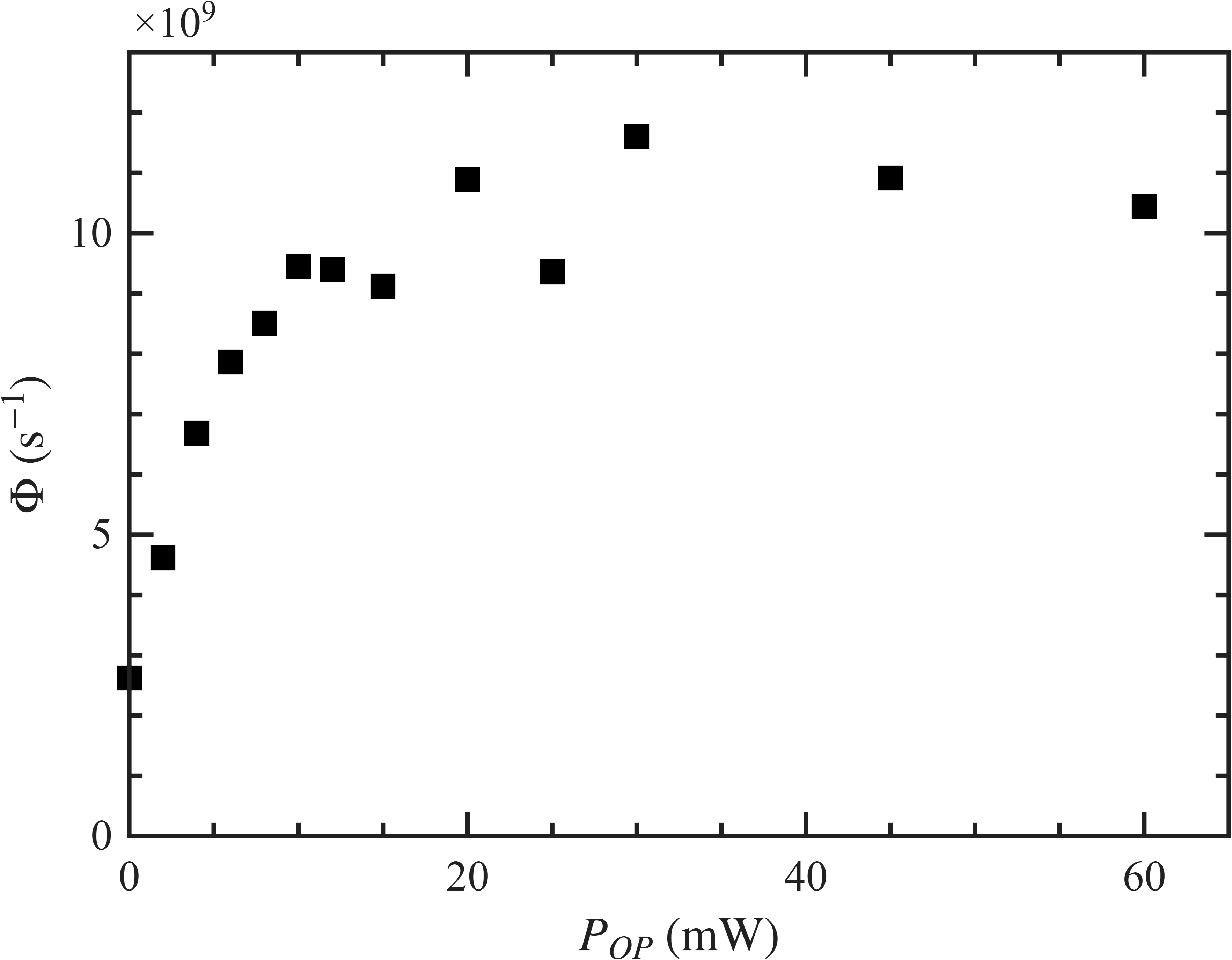}
\caption{\label{fig:fluxVsOP} The flux of slow atoms $\Phi$ as a function of OP laser power.}
\end{figure}

Fig.~\ref{fig:fluxVsOP} shows the influence of optical pumping on the flux of slow atoms. 
If a uniform distribution of atoms is assumed among all energy levels of the hyperfine split ground state, the maximum enhancement is limited to a factor of $8$. 
Overall, the measured flux increases by more than a factor of $4$ if a high power optical pumping is applied. 
Although we believe this value can be further improved to some extent by optimizing the quantization field value and beam position, it does show a significant enhancement in performance.
We conclude that optical pumping is successful in transferring a large portion of the population to the target energy level, which is subject to a consequent deceleration.

\section{Conclusions}
In conclusion, we design and build a Zeeman slower for $^7$Li atoms on the basis of standard permanent bar magnets.
The magnets are set in Halbach configuration, allowing for the creation of a uniform magnetic field in the transverse direction.
Although the latter dictates the loss of $50\%$ of the optical power of the slowing beam, it allows for a single slowing laser frequency for the entire deceleration region and a clearly defined and separated region for optical pumping prior to deceleration. 

The structure designed to support the magnets mechanically is created using a 3D printer and a standard material (PLA). 
We demonstrate that the Zeeman slower is efficient and obtain a high flux of clod atoms in the region of magneto-optical trap.

This design can be easily adopted to use with other atomic species in all possible realizations of the Zeeman slower such as increasing, decreasing, and zero-crossing magnetic field configurations.

\section{Acknowledgments}
This research was supported in part by the Israel Science Foundation (Grant No. 2284/24) and by a grant from the United States-Israel Binational Science Foundation (BSF, Grant No.~2022740).

% \bibliographystyle{aipnum4-1}
%\bibliography{Zeeman}  % or whatever your .bib file is called


\end{document}